\documentclass[aps,prl,twocolumn,groupedaddress,showpacs]{revtex4-1}
\usepackage{graphicx}
\usepackage{amsfonts}		
\usepackage{color}
\usepackage{multirow}

\def\ie{\hbox{\it i.e.}}
\def\beq{\begin{equation}}
\def\eeq{\end{equation}}
\def\bea{\begin{eqnarray}}
\def\eea{\end{eqnarray}}

\pacs{75.10.Nr, 75.50.Lk}

\begin{document}

\title{Diluted Antiferromagnetic 3D Ising model in a field}
\author{M.~Picco$^{1,2}$ and N.~Sourlas$^3$}

\affiliation{$^1$Sorbonne Universit\'es, UPMC Univ Paris 06, UMR 7589, LPTHE, F-75005, Paris, France}
\affiliation{$^2$CNRS, UMR 7589, LPTHE, F-75005, Paris, France}
\affiliation{$^3$Laboratoire de Physique Th\'eorique de l'Ecole Normale Sup\'erieure, 
CNRS and UniversitŽ Pierre and Marie Curie,
75005 Paris, France}
\date{\today}

\begin{abstract}
\noindent
We present numerical simulations for the diluted antiferromagnetic 3D Ising model (DAFF) in an external 
magnetic field at zero temperature. 
Our results are compatible with the DAFF being in the same universality class as 
the Random Field Ising model, in agreement with the renormalization group prediction. 
\end{abstract}

\maketitle

Despite numerous efforts phase transitions in disordered systems are not yet 
fully understood. In experiments it is difficult to reach thermal equilibrium. 
The same is true in many numerical simulations which face the additional difficulty 
of very large sample to sample fluctuations. 
It turns out that the random field Ising model (RFIM) is the only case where 
this difficulty can be overcome due to the possibility of finding exact ground states (no 
thermalisation problem), using a very fast algorithm \cite{O,AdA}. This allowed 
the simulation of very large systems with high statistics \cite{O,AAS,MF,FMM}.
On the theoretical side, the RFIM and the diluted branched polymers 
are the only models where renormalization group can be 
carried out to all orders of perturbation theory \cite{young,ps1,ps2}. 
It predicts dimensional reduction in both cases. 
Here dimensional reduction means that the critical exponents of the RFIM in $D$ dimensions are the same 
with the exponents of the ferromagnetic Ising model in $D-2$ dimensions.
Dimensional reduction has been proven true for the diluted branched polymers \cite{BI}, while it is not true 
for the RFIM \cite{BK}.

One of the most striking predictions of perturbative renormalization group (PRG) is that the RFIM 
and the diluted antiferromagnets in an external magnetic field (DAFF)  are in the 
same universality class \cite{FA,Cardy}.
This is very important because it allows the connection with experiments: 
there are no experimental realizations of the RFIM, while the DAFF has been studied 
experimentally extensively \cite{Belanger1,SBFB,Belanger2}. 
PRG universality predicts that critical exponents and other universal quantities (see later) 
take the same values for the RFIM and the DAFF 
 and that they do no depend on the random field probability 
distribution for the RFIM or the dilution for the DAFF. 
The validity of this prediction is not guaranted because other predictions of the PRG (dimensional reduction) are false. 
Indeed it has been speculated  that the RFIM and the DAFF are not in the same universality class \cite{sou,AXHK}. 

In the present paper we establish numerically for the first time that the RFIM and the DAFF are 
in the same universality class and that 
universal quantities do not depend on the dilution probability in the DAFF as predicted by the PRG. 
It has ben shown recently that in the case of the RFIM, 
different random field probability distributions of the random field 
belong to the same universality class \cite{FMM}. 
No clear picture has yet emerged from the experimental studies.
These results raise two questions: why some predictions of 
PRG are verified, while others are not, and why the critical exponents 
measured experimentally \cite{Belanger1,SBFB,Belanger2} are different for different experiments and from 
the values obtained by numerical simulations.

A very large number of numerical simulations have been devoted to the study of the RFIM. 
Much less effort has been devoted to the numerical study of the DAFF \cite{sou,FMMY,AXHK}.
 
We start from the Ising antiferromagnet in a field~:
$$
{\cal H}_{AF} =  J \sum_{<ij>} \sigma_i \sigma_j - H_0 \sum_{i}  \sigma_i \; ,$$
with the spins $\sigma_i = \pm 1$ on a cubic lattice of linear size $L$. We considered nearest neighbour interactions 
and periodic boundary conditions. 
$H_0$ is an external constant magnetic field. The coupling between spins is antiferromagnetic,
\ie\ $J > 0$. 
It is convenient to perform a gauge transformation 
$ \sigma(x,y,z) \to (-1)^{x+y+z} \sigma(x,y,z) $.
For a cubic lattice, periodic boundary conditions and even $L$ we obtain, 
because of the absence of frustration,   a ferromagnet in an alternating magnetic field
$$
{\cal H}_{AF} =  -J \sum_{<ij>} \sigma_i \sigma_j - H_0 \sum_{i}  (-1)^{i_x+i_y+i_z} \sigma_i \; .
$$
We will consider this model in the presence of random site dilution. For this  
we replace the spin variables $\sigma_i$ with $\epsilon_i \sigma_i$ where $\epsilon_i$ are 
independent quenched random variables which take the value 
$0$ with a probability $d$ (dilution) or $1$ with a probability $1-d$. 
Then the Hamiltonian for a given configuration of dilution 
$\epsilon_i$ (instance) is 
\beq
\label{HDAFF}
{\cal H}_{DAFF} =  -J \sum_{<ij>} \epsilon_i \epsilon_j \sigma_i \sigma_j - H_0 \sum_{i}  (-1)^{i_x+i_y+i_z} 
\epsilon_i \sigma_i \; . 
\eeq

The relevant parameters of the DAFF are the temperature $T$ and the ratio $R= H_0/J $.
The phase diagram of the DAFF is a line in the $ T - R$ plane. 
It is believed that all points on this line belong to the same universality class.
This line crosses the $T=0$ axis, \ie\ there is a phase transition at $T=0$.
It is well known that it is possible to find exact ground states of the 
ferromagnetic RFIM  using a very fast optimisation algorithm \cite{O,AdA}. 
The properties of the phase transition of the RFIM have been studied by the extensive use 
of this algorithm \cite{O,AdA,AAS,MF,sou}.

The goal of transforming the original Hamiltonian to the one of eq.~(\ref{HDAFF}) is 
to allow the use of this algorithm \cite{sou}. In this paper we present the results 
of extensive numerical simulations of the DAFF on a cubic lattice in three dimensions 
with periodic boundary conditions. 
We have studied the case of dilution probabilities, $d=0,05$, $d=0.07$ and $d=0.37$. 
Our results are compatible with universality for all three values of $d$. The renormalization group 
argument for universality is valid for small $d$, $d< d_c$, but there is no estimation of $d_c$. 
It has been speculated that $d_c$ may be the percolation threshold  $d_p=0.3116077(4)$ \cite{DB}.
We find universality even for $d=0.37$ $> d_p $.

In order to locate the phase transition and extract the critical properties, we compute 
the ground state magnetization for different values of $R$ and sizes $L$ for a large number of 
samples, typically $10^6$. 
From the magnetization we compute the dimensionless Binder like magnetic cumulant $ U_4 $ 
\beq
U_4(R,L) = {[m(R,L)^4] \over [m(R,L)^2]^2} \; ,
\eeq
where $[A]$ is the average of $A $ over the samples. 
We also compute the dimensionless ratio $\xi(R,L)/L$, where $\xi$ is the  correlation length 
and $L$ the system linear size. The correlation length is defined from the wave-vector susceptibility 
$\chi(\vec{k}) = \left[ <  (\sum_j \sigma_j e^{i \vec{k} \cdot \vec{r}_j })^2>\right]/ N^2$
as \cite{CoFrPr} : 
\beq
\xi = {1\over 2 \sin(\pi/L)} \sqrt{{\chi(0)\over \chi(\vec{k}_{min})}-1}\; ,
\eeq
with $N$ the number of spins and $\vec{k}_{min} = ({2\pi\over L},0,0)$. 

We compute the $L$ dependant effective critical values of $R$, $R_{c,U} (L)$  and $R_{c,\xi } (L) $. 
$R_{c,U} (L)$ is the value of $R$ for which $U_4(R,L) = U_4(R,2 L) $, \ie\ the value of $R$ 
at which $U_4(R,L)$ and $U_4(R,2 L) $ cross and $U_4^c(L) $ the value of $U_4$ at the crossing. 
Similarly $R = R_{c,\xi } (L) $ is the crossing point of $\xi(R,L)/L$, \ie\ $\xi(R,L)/L =\xi(R,2L)/2L$ 
and $\xi^c(L)/L $ the value of $\xi(L)/L$  at the crossing. 
According to the renormalization group, for large $ L $, 
 $R_{c, U} (L) $ and $   R_{c,\xi  } (L) $ should converge to the same 
  value $ R^c $ which is the critical value of $R$. Finite size scaling implies 
\beq    
\label{eq3}
R_{c, U} (L) \to R^c + a_U L^{- 1/\nu -\omega}  \; ; \; 
 R_{c,\xi } (L)  \to  R^c + a_\xi L^{- 1/\nu -\omega} 
 \eeq
 where $ \nu $ is the correlation length exponent and $ \omega $ the exponent of the first non 
 leading correction to scaling. 
 
Similarly $ U_4^c(L)  \to U_4^c + b_U L^{ -\omega} \; ;  \; \xi^c(L)/L  \to \xi^c/L  +  b_{\xi} L^{ -\omega}$.
$U_4^c $, $\xi^c/L$ and the exponents $\nu$ and $\omega$ are universal quantities.

We will show now the existence of discontinuities of 
the ground state magnetization as a function of the ratio $R$.  
These discontinuities will affect the determination of the crossing points, mentioned above.
We illustrate these discontinuities in the case of $d=0.07$.
\begin{figure}[h]
\begin{center}
\includegraphics[scale=1.2]{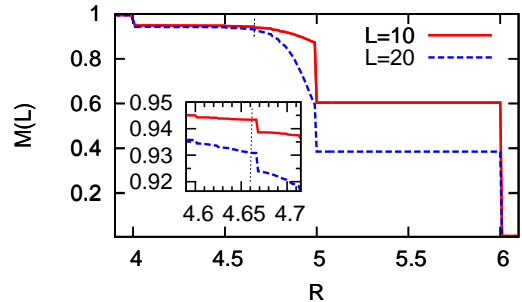}
\end{center}
\vskip -0.5cm
\caption{(Color online.) Magnetization vs. $R$ for the 3D DAFF at $T=0$ with dilution $d=0.07$. 
The inset contains an expansion of the neighbourhood of the critical point.
}
\label{Pmag}
\end{figure}
\begin{figure}[h!]
\begin{center}
\hskip -0.4cm
\includegraphics[scale=0.94]{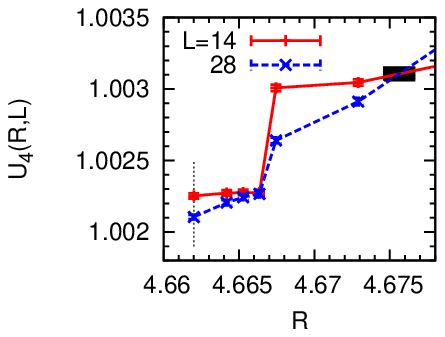}
\hskip -0.8cm
\includegraphics[scale=0.94]{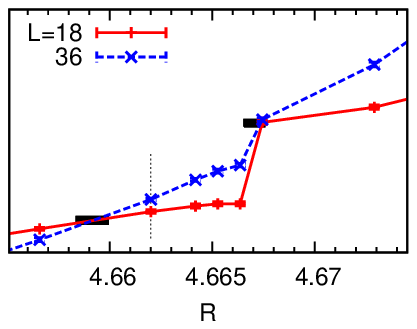}
\end{center}
\vskip -0.5cm
\caption{(Color online.) Crossing of $U_4$ versus $R$ for the 3D DAFF 
at $T=0$ with $d=0.07$ for $L-2L=14-28$ and $18-36$.
The vertical dashed line corresponds to the critical $R^c \simeq 4.662$. 
}
\label{PJ}
\end{figure}
Large discontinuities of the magnetization were already observed in the first simulations of the DAFF \cite{sou}. 
They also occur in the 3D RFIM when the probability distribution of the random field is bimodal \cite{AAS}.
In both cases,  the position of these discontinuities is 
independent of the  system size $L$ as illustrated below, \ie\ they don't behave as in eq.(\ref{eq3}). 
They occur when 
the ratio $R=H_0/J$ goes through a rational number. 
Because of the dilution, there are spin clusters weakly connected to the rest 
of the lattice. By changing the value of $R$ it becomes energetically favourable 
to flip these clusters of spins. This is an artefact of zero temperature. 
Figure~\ref{Pmag} shows the average magnetization 
as a function of $R$, for $L=10$ and $L=20$. We observe several such discontinuities at values of $R$ which are  
size independent. 
The inset of the Figure show the magnetization close to the critical value $R^c 
\simeq 4.662$ (this value will be determined below). Note that it is 
very close to a large jump occurring 
at $ R = 14/3 \simeq 4.6666 $. This jump corresponds to a change of ground state by 
 flipping clusters of five spins, four spins $+$ and one spin $-$, $\Delta M = 6 $ 
with a breaking of $14$ bonds. The jump of the magnetization will also affect other quantities, in particular  
$U_4(R,L)$.

\begin{figure}
\begin{center}
\vskip -0.1cm
\hskip -0.3cm
\includegraphics[height=3.5cm,width=4.5cm]{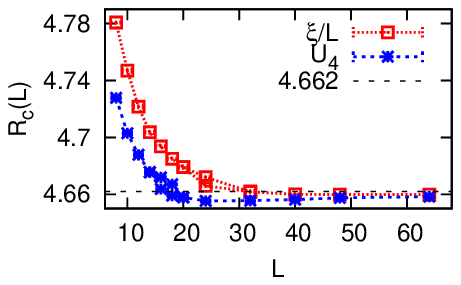}
\hskip -0.3cm
\includegraphics[height=3.5cm,width=4.5cm]{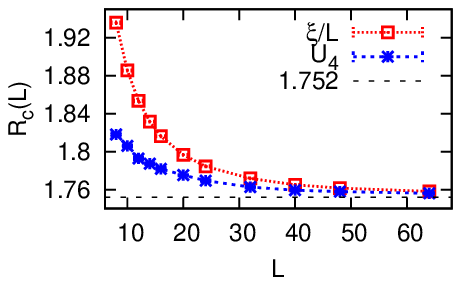}
\end{center}
\vskip -0.5cm
\caption{(Color online.) $R_c(L)$ as a function of $L$ for the correlation length $\xi /L$ and $U_4$. 
The left part corresponds to the dilution $d=0.07$ 
and the right part to $d=0.37$.}
\label{Crossing}
\end{figure}

\begin{figure}[h!]
\begin{center}
\hskip -0.3cm
\includegraphics[height=3.6cm,width=4.5cm]{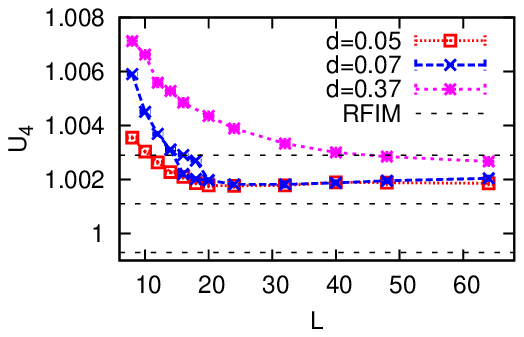}
\hskip -0.3cm
\includegraphics[height=3.6cm,width=4.5cm]{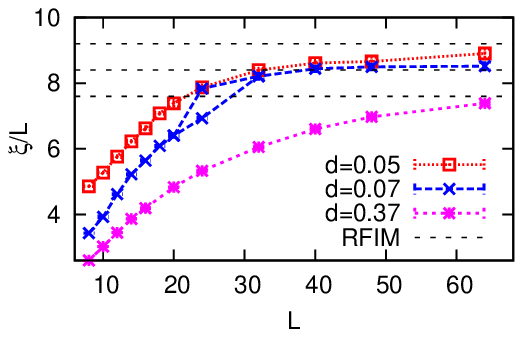}
\end{center}
\vskip -0.5cm
\caption{(Color online.) Crossing of $U_4$ (left part) and of $\xi/L$ (right part) 
for the AF 3D IM with dilution $d=0.05, 0.07, 0.37$ and $T=0$. 
}
\label{PU4}
\end{figure}

In Figure \ref{PJ}, we show the crossings of $U_4(R,L)$ versus $R$ for two pairs of sizes $L - 2L$. 
In each case we observe again a jump of $U_4(R,L)$ for $R=14/3$.
Since this value is very close to the critical value for $R$, it will  affect the 
measurements. In particular there can be more than one 
crossings of $U_4(R,L)$ and $U_4(R,2L)$.
In Figs.~\ref{Crossing}-\ref{PU4}, we  show both crossings. 

In order to get rid of those spurious singularities a gaussian random component of small amplitude $dh_i $ was 
added to the external field $H_0$ \cite{sou}, \ie\
$$ H_i = H_0  (-1)^{i_x+i_y+i_z} + dh_i, \ \ \ dh_i= w h_i    \; .$$
$ h_i $ are independent gaussian random variables of mean zero and variance one and $ w $ 
is the strength of this additional quenched disorder. 
In both cases of the RFIM with bimodal field distribution and the DAFF 
the size independent spurious singularities disappear with the addition of $dh_i $. 
We will now argue on renormalization group arguments, that in the case of the DAFF, this 
 additional disorder may change the universality class. 
First consider the Ising antiferromagnet in a field without any dilution or disorder. 
The Landau-Ginsburg-Wilson Hamiltonian is 
\bea
{\cal H} &=&  -J \sum_{<xy>}  \phi(\vec {x}) \phi(\vec {y}) \\ 
&&- H_0 \sum_{\vec {x}}  (-1)^{i_x+i_y+i_z}  \phi ( \vec {x})  -g  \sum_{\vec {x}}  \phi(\vec {x})^4 \nonumber \; .
\eea
We separate even and odd sites by defining 
$ \psi_{\pm} (\vec {x}) =  \phi(\vec {x}) (1 \pm (-1)^{i_x+i_y+i_z} )/2 $. In terms of the new variables 
$ \sum_{\vec {x}}  (-1)^{i_x+i_y+i_z}  \phi ( \vec {x}) =  \sum_{\vec {x}} ( \psi_{+} (\vec {x})  
 -\psi_{-} (\vec {x})) $ and
\bea
{\cal H} &=&  -J \sum_{<xy>} (\psi_{+} (\vec {x}) \psi_{-} (\vec {y}) + 
\psi_{-} (\vec {x}) \psi_{+} (\vec {y}))  \nonumber \\
&& - H_0  \sum_{\vec {x}} ( \psi_{+} (\vec {x})   -\psi_{-} (\vec {x}))
 \\
&& - g (\psi_{+}^4(x) + \psi_-^4(x) + 2 (\psi_+^2(x) \psi_-^2(x)) )  \nonumber \; . 
\eea
 This Hamiltonian has the symmetry $ \psi_{+} (\vec {x}) \to - \psi_{-} (\vec {x}) $ and 
 $ \psi_{-} (\vec {x}) \to - \psi_{+} (\vec {x}) $, \ie\
 $ \phi(\vec {x}) \to - \phi(\vec {x}) $. At the phase transition this symmetry is 
 broken with the appearance of a spontaneous magnetization, while the field 
 $  \psi_{+} (\vec {x})  - \psi_{-} (\vec {x})  $ is non critical and decouples. In other words the 
 antiferromagnet in a constant field in a cubic lattice belongs to the universality class of the 
 ferromagnetic Ising model. By adding a quenched random component to $H_0$ we change 
 universality class to the one of the RFIM. This is true in the absence of any dilution. 
We have verified numerically that this is true.

In this paper we consider both cases of $w=0.1 H_0 $ and $w=0.0$.
We found that both models are in the same universality class as the RFIM. 

We first present our results when a small random field is added to $H_0$. 
\begin{figure}[h]
\begin{center}
\includegraphics[height=4.0cm,width=8.0cm]{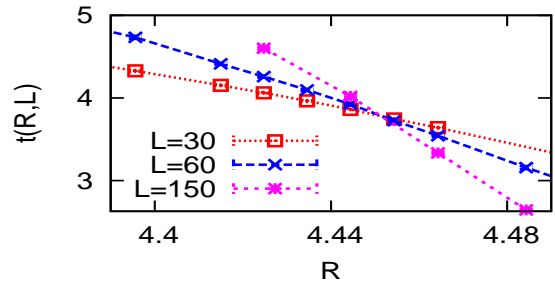}
\end{center}
\vskip -0.5cm
\caption{(Color online.) $t(R,L)$ vs. $R$ for the DAFF with dilution $d=0.07$ 
and $T=0$ in a noisy external field.
}
\label{PTN}
\end{figure}
We studied the case of $ d = 0.07$, $H_0=1.0$ and $ w = 0.1$.
For every sample we compute the ground state magnetization $m$ for different 
values of the ratio $R$.
We compute the magnetic cumulant $U_4$ defined above and  in order to take into account the constrain 
$ 1 \le U_4(R,L) \le 3 $ we change variables $ x(R,L) = ( 3 - U_4(R,L) ) / 2 $, 
$ 0 \le x(R,L) \le 1 $ and $ \tanh{(t(R,L))} =  x(R,L)  $. 
In Figure~\ref{PTN} we show $t(R,L)$ as a function of $R$ and $L \ge 30$. 
We find that the data are compatible with the finite size scaling hypothesis 
$t(R,L) = F((R-R^c) L^{1/\nu})$, without needing subdominant corrections in $ L $.
We fit the data with the ansatz
\bea
t(R,L) &=& F((R-R^c) L^{1/\nu}) = t_0 + t_1 (R-R^c) L^{1/\nu} \nonumber \\
&&  + t_2 (R-R^c)^2 L^{2/\nu} + t_3 (R-R^c)^3 L^{3/\nu} \; , \nonumber \
\eea
valid for $R \simeq R^c$. 
We found $R^c = 4.4516 (50)$, $ \nu =1.43 (14) $, $ U_4  =1.0021 (9) $
in excellent agreement with Fytas and Martin-Mayor \cite{FMM} who found for the RFIM 
$ U_4 = 1.0011 (18) $ and $ \nu =1.38 (13) $. 
We conclude that our data are compatible with the statement that the DAFF with $ H_i = H_0+ dh_i $ is  
in the same universality class with the RFIM, \ie\ 
the addition of dilution does not change the universality class in this case.

Next we present our results when $dh_i = 0 $ , \ie\ constant external field with no 
addition of a random component. For different sizes $L$ and different values of $R$ 
we compute the following dimensionless ratios 
$X(R,L)$. $X(R,L)$ is  $U_4(R,L)$ or $\xi(R,L) / L$.
These are the same quantities considered in \cite{FMM} and it 
will allow a direct comparison of our results with those of the RFIM.
In order to determine the critical value of $R^c$, we compute the crossing value $R_X (L) $ of $R$ 
for which $ X(2L) = X(L) = X_c (L) $ and the value $ X_c (L) $ at the crossing as explained above. 
Renormalization group predicts that 
$ \lim_{L \to \infty } X_c (L) $ is universal.

The results of the crossing for increasing linear sizes $L$ are shown in Figure~\ref{Crossing} 
for the values of dilution, 
$d=0.07$ and $0.37$. The lines are linear extrapolations between crossings. 
We observe two crossing points for some values of $L$, resulting in the  doubling of some sections of the 
lines in the graph. This is due to the discontinuities of the magnetization discussed above. 
In both cases, the convergence is very fast. 
The doubling generates a small uncertainty in the extrapolation to $L \to \infty $ which  
is taken into account in the error bars.  We see that $R_{U_4} - R_{\xi/L}$ converges to zero very fast with $L$.
Non leading corrections to the large volume limit are negligible and we can easily extrapolate to $L \to \infty $. 
 We find $R^c = 4.662 (1)$ 
for $d=0.07$ and $R^c=1.752 (2)$ for $d=0.37$ and $R^c = 4.8875(10)$ for $d=0.05$ (not shown here). 
 
Figure \ref{PU4} shows the values of  $U_4$ and $\xi/L$  at the crossings for increasing sizes $L$. 
We observe that each of these quantities  converges nicely toward their asymptotic limit. 
These limits are compatible to be dilution independent.  
We determined the asymptotic values $U_4 = 1.0020 (5)$ and $\xi/L = 8.5 (5)$. 
The convergence is slower for the dilution probability 
$ d=0.37$. The existence of multiple crossing points is again visible in the plots but affects very 
little the  asymptotic values, which are fully compatible with those of  the 3d RFIM \cite{FMM} and this for all the three 
dilution considered. 
The values for the RFIM are shown in the figures as dashed lines. 
The middle one corresponds to the value determined in  \cite{FMM}  
and the two other ones show the upper and lower error bars. 

Perturbative renormalization group correctly describes universality classes. 
The agreement is very good for $U_4$ and $\xi/L$. 
%

These results make us confident that the DAFF  belongs to the same universality class with the RFIM. 
We assume that this is the case and that 
the magnetic susceptibility exponent $\gamma$ and $\nu $ take their RFIM values \cite{FMM,PicSo}
$ \gamma / \nu = 1.48$  and $ 1/\nu = 0.7 $. 
As in \cite{PicSo}, we have applied small additional translation invariant magnetic fields 
$\delta h_k $, $k=1,2, \cdots$ and change the ferromagnetic coupling to $j_c - dj$. 
$dm_k$ is the variation of the ground state magnetization $m_k$ due to $\delta h_k$ and $dj$. 
We have computed the probability distribution $ P(dm, dj, \delta h,L) $ of $ dm_k$ for 
different values of $dj, \delta h$ and $L$. 
\begin{figure}[h!]
\begin{center}
\vskip -0.2cm
\includegraphics[height=4.0cm,width=9.0cm]{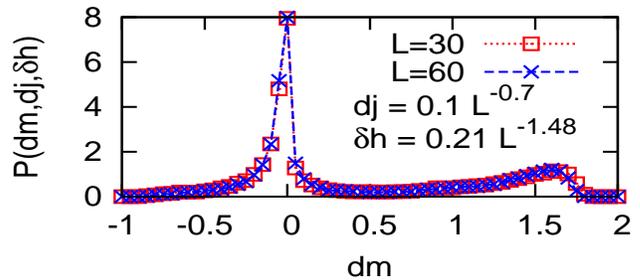}
\end{center}
\vskip -0.5cm
\caption{(Color online.)  $P(dm, d j , \delta h )$ vs. $dm$ for $L=30$ and $60$ for $p=0.07$ and $w=0.1$.
}
\label{Pdm1}
\end{figure}
In Fig~\ref{Pdm1} we show  $P(dm, dj, \delta h,L) $ where we simultaneously scale $dj = 0.1 L^{-0.7}$ 
and $\delta h = 0.21 L^{-1.48}$.
We observe that this probability distribution indicates a strong violation of self averaging and obeys 
a perfect finite size scaling. 
This is exactly what happens in the case of the RFIM \cite{PicSo} and with the same values of the critical 
exponents. It is consistent with the hypothesis 
that the exponents $\gamma$ and $\nu$ take the same values as in the RFIM, confirming again universality. 

In this paper we study the critical behaviour of the diluted antiferromagnet in 
a field in three dimensions. 
Our results are fully compatible with the prediction of the perturbative renormalization group (PRG) 
that it belongs to the same universality class with the random field Ising model. 
The PRG prediction is for small dilution. It is quite remarkable that a dilution as small as $d=0.05 $ 
changes the critical values so much, 
in agreement with PRG. 
We remind that for the 3D Ising ferromagnet \cite{H} $U_4=1.60361(1) $ and $\nu = 0.63002(10)$.
We found that universal quantities do not depend on $d$ up to the largest value of $d$ 
we studied, $d=0.37 $. 

There are two very important points we still do not understand~:
i) Why the perturbative renormalization group predicts correctly universality classes, 
a highly non trivial prediction, while it fails so much in the prediction of 
critical exponents (no dimensional reduction). 
ii) Why numerical simulations do no not agree with the 
experimental results \cite{Belanger1,SBFB,Belanger2}. 
Is this due to the difficulty to equilibrate experimental samples ?

Further studies are required to elucidate these two points.

\end{document}